\documentclass[prl,aps,floatfix,twocolumn,showpacs]{revtex4}
\usepackage{epsf}
\usepackage{mathptmx}
\usepackage{amsmath}
\usepackage{amssymb}
\usepackage{amsfonts}
\usepackage{bm}
\newcommand{\Eq}[1]{Eq.\! (\ref{#1})}
\newcommand{\Eqs}[1]{Eqs.\! (\ref{#1})}
\begin{document}

\title{Current states in superconducting films: numerical results and
approximations}

\author{E.V.~Bezuglyi}
\email{eugene.bezuglyi@gmail.com}
\affiliation{B.Verkin Institute for Low Temperature Physics and Engineering,
Kharkov 61103, Ukraine}

\begin{abstract}
We present numerical solution of equations by Aslamazov and Lempitskiy (AL)
\cite{AL} for the distribution of the transport current density in thin
superconducting films in the absence of external magnetic field, in both the
Meissner and the vortex states. This solution describes smooth transition
between the regimes of a wide film and a narrow channel and enables us to find
the critical currents and current-voltage characteristics within a wide range
of the film width and temperatures. We propose simple approximating formulas
for the current density distributions and critical currents.
\end{abstract}\vskip -3mm


\maketitle

\section{I. Introduction} \vskip -3mm

The main property of the current states in wide supercon\-duc\-ting films,
which distinguishes them from narrow channels, is an inhomogeneous distribution
of the current density $j$ due to the Meissner screening of the current-induced
magnetic field. It should be noted that the current state of a wide thin film
is qualitati\-vely different from that in a bulk supercon\-duc\-tor. Whereas
the transport current $I$ in the latter case flows only within a thin surface
layer with the thickness of the order of the London penetration depth
$\lambda$, the current in a thin film with the thickness $d \ll \lambda$ is
distributed over its width $w$ according to the approximate power-like law $j
\sim [(w/2)^2-x^2]^{-1/2}$ \cite{LO,AL}, where $x$ is the transversal
coordinate with the origin in the middle of the film. Thus, the characteristic
length $\lambda_\perp(T)=2\lambda^2(T)/d$, which is commonly referred to as the
penetration depth of the perpendicular magnetic field, has nothing to do with
the scale of the current decay, but rather plays the role of a ``cutoff
factor'' in the above-mentioned law of the current distribution at the
distances $\lambda_\perp$ from the film edges and thereby determines the
magnitude of the edge current density. The latter was estimated in \cite{LO} as
$j_e \approx I/d\sqrt{\pi w\lambda_\perp}$, assuming $w$ to be much larger than
$\lambda_\perp(T)$ and the coherence length $\xi(T)$.

In such an inhomogeneous situation, the resistive transition of a wide film
occurs \cite{LO,AL,Shmidt,Likharev} when $j_e$ reaches the value close to the
critical current density $j_c^{GL}$ in the Ginzburg-Landau (GL) theory. Using
this estimate, the expression $I_{c} \approx j_{c}^{GL}d\sqrt{\pi
w\lambda_\perp}$ for the critical current has been obtained in \cite{LO}. This
equation is widely used in analysis of experimental data (see, e.g.,
\cite{Andratskiy,Musienko}) and imposes a linear temperature dependence of the
critical current $I_{c}(T) \propto 1-T/T_{c}$ near the critical temperature
$T_{c}$. A quantitative theory by Aslamazov and Lempitskiy (AL) \cite{AL} also
predicts the linear dependence $I_{c}(T)$ but gives its magnitude numerically
larger than the above estimate. This result has been confirmed in recent
experiments \cite{Dmitriev,Dm1,Zol1}.

The instability of the current state at $I=I_c$ results in the entry of
vortices whose motion leads to the appearance of the electric field, i.e., to
formation of the vortex part of the $I$-$V$ characteristic (IVC). While the
current further increases, the motion and annihilation of the vortices form a
peak in the current density at the center of the film. For certain current
value $I_m$, the magnitude of this peak reaches $j_c^ {GL}$, which results in
instability of the stationary flow of the vortices \cite{AL}. Further behavior
of the film depends on the conditions of the heat removal \cite{Kaplan} and the
quality of the films. In experiments performed decades ago, an abrupt
transition to the normal state has been usually observed at $I=I_m$, whereas in
later researches, in which optimal heat compliance was provided, a step-like
structure of the IVC is observed at $I>I_m$ (see, e.g.,
\cite{9,10,11,Dmitriev,13}). This indicates the appearance of phase-slip lines,
similar to the phase-slip centers in narrow channels.

Since $\lambda_\perp(T)$ unlimitedly grows at $T\to T_{c}$, any film reveals
the features of a narrow channel in the immediate vicinity of $T_{\rm c}$: at
$\lambda_\bot \gg w$, its critical current is due to the uniform pair-breaking
(narrow channel regime) thus showing the temperature dependence of the GL
pair-breaking current $I_{c}^{GL}(T) \propto (1-T/T_{c})^{3/2}$. As the
temperature decreases, the film exhibits a crossover to an inhomogeneous
current state, in which vortex nucleation is responsible for the resistive
transition (wide film regime). In the experiment \cite{Dm1,Zol1}, the linear
temperature dependence of $I_c$ predicted in \cite{LO,AL} for wide films
becomes pronounced only at low enough temperatures, when $\lambda_\perp(T)$
becomes smaller than the film width by the factor of $20-30$, although the
vortex state already occurs at much larger values of $\lambda_\perp \sim w/4$.
Similar difficulties in the fitting of the IVC measurements with the
asymptotical results of the AL theory were met in the experiment \cite{Zol2},
because the condition $I_c \ll I_m$, used in \cite{AL}, can be fulfilled only
in extremely wide films whose width exceeds $\lambda_\perp(T)$ by several
orders of magnitude. Thus, there exist a considerable intermediate region of
the film widths and temperatures, where the asymptotic results of the AL theory
cannot give a satisfactory description of the experimental data, although the
assumptions and initial equations of this theory remain valid in this region.

In order to fill up this gap, we perform in this paper a numerical solution of
the AL equations within a wide region of the ratio $w/\lambda_\perp$. The
results of our computations describe smooth transition between the regimes of a
wide film and a narrow channel and demonstrate evolution of the current density
distribution with the increase of the transport current in both the Meissner
and the vortex states. We notice that within the AL theory, the critical
currents $I_c$ and $I_m$, being normalized on the GL critical current
$I_{c}^{GL}$, as well as the specifically normalized IVC, are universal
functions of the ratio $w/\lambda_\perp(T)$. We calculate the fitting constants
in the asymptotic formulas of the AL theory and propose approximating
expressions for the current density distributions, which are in rather good
agreement with the results of numerical computations.

\section{II. Basic equations and results of the AL theory}\vskip -3mm

A starting point of the AL theory are the static GL equations for the
dimensionless modulus $F$ of the order parameter (normalized on its equilibrium
value in the GL theory) and the gauge-invariant vector potential ${\bf Q} =
{\bf A} - \kappa_{\textrm{eff}}^{-1}\nabla \chi$,
\begin{align} \label{GLeq1}
&\kappa_{\textrm{eff}}^{-2} \nabla^2 F + F(1-F^2- {\bf Q}^2)=0, \\
&{\textrm {rot rot}} \,{\bf Q} = -F^2 {\bf Q} \delta(z).\label{GLeq2}
\end{align}
Here ${\bf A}$ is measured in units of $\Phi_0/2\pi\xi$, $\Phi_0$ is the
mag\-netic flux quantum, $\chi$ is the order parameter phase and
$\kappa_{\textrm{eff}} = \lambda_\bot/\xi$ is the effective GL parameter. The
axis $z$ is directed perpendicular to the film whose thickness is assumed to be
infinitely small, and all distances are measured in units of $\lambda_\bot$.

Usually in thin films, the GL parameter is large, $\kappa_{\textrm{eff}} \gg
1$. Assuming the film width much larger than $\xi(T)$, one thus can neglect the
gradient term in \Eq{GLeq1} and use the local relation $F^2 = 1 -{\bf Q}^2$
between the order parameter and the vector potential. Inside the thin film, the
latter has only one component $Q \equiv Q_y$ and can be found from equation
\begin{align} \label{EqQ}
\frac{dQ}{dx} = -\frac{1}{2\pi}\int_{-\widetilde{w}/2}^{\widetilde{w}/2}
\frac{Q(x')[1-Q^2(x')]}{x'-x}dx',\quad \widetilde{w}=w/\lambda_\bot,
\end{align}
with the Biot-Savard integral which relates the magnetic field $dQ/dx$ to the
dimensionless density $j=Q(1-Q^2)$ of the surface current. Equations
\eqref{GLeq1}--\eqref{EqQ} determine the stability threshold of the Meissner
state, when the vortices begin to penetrate into the film and the edge value of
the vector potential appears to be close to its critical value
$Q_c^{GL}=1/\sqrt{3}$ in the GL theory for narrow channels; this fact will be
used in our calculations \cite{footnote}. The asymptotic value of the critical
current at $w \gg \lambda_\perp$ was calculated in \cite{AL} and then refined
in \cite{BL}:
\begin{equation}
\label{eq6} I_c^{AL} = \sqrt{15/8}\, I_c^{GL} \left( \pi \lambda_\bot/w\right)
^{1/2}.
\end{equation}

According to \cite{LO,Ivan,AL}, the resistive vortex state of a wide film can
be described within a hydrodynamic approximation for the viscous motion of the
vortex fluid, by including the contribution of the vortices $n\Phi_0$ [$n(x)$
is the vortex density] to the net magnetic field induction. Then, using the
continuity equation for the flux density $nv$ of the vortex fluid, expressing
the vortex velocity $v$ through the linear current density $j$ and the
viscosity coefficient $\eta$ as \cite{Gor}
\begin{equation}
\label{v} v = - \eta^{-1}\Phi_0 j \,\textrm{sign} \,x,
\end{equation}
and the average electric field -- through the flux density as $E = -nv\Phi_0$,
the authors of \cite{LO,Ivan,AL} arrive at the equation
\begin{equation}
\label{eq1} 4\pi \frac{\lambda_\bot}{w}\frac{dj}{dx} + 2\,V.p.\int_{-1}^{1}
\frac{j(x')dx'}{x' - x} = - \frac{\eta c^3 E}{\Phi_0 j(x)}\,\textrm{sign} \,x
\end{equation}
(a mismatch of the coefficient in the first term with that in \cite{LO,AL} is
due to the difference in the definition of $\lambda_\bot$). Here and below, the
coordinate $x$ is normalized on the film half-width $w/2$, and the expression
$\textrm{sign} \,x$ indicates the opposite direction of the vortex motion in
different halves of the film.

An asymptotic analysis of \Eq{eq1} at $w \gg \lambda_\perp$ shows \cite{AL}
that the IVC is linear in the vicinity of $I_c$, whereas at large currents, the
voltage grows quadratically,
\begin{equation}
\label{AL_IVC} V = E_0 L\begin{cases} (I-I_c)/I_c, & I-I_c \ll I_c; \\
C(I/I_c)^2, & I \gg I_c, \end{cases} \quad E_0 = \frac{8\Phi_0 I_c^2}{\eta w^2
c^3}.
\end{equation}
The current distribution in the middle of the film has the peak of the order of
${\ln^{1/2}(w/\lambda_\perp)}$ which leads to the following estimate of the
maximum current of existence of the vortex state,
\begin{equation}
\label{AL_im} I_m = C' I_c^{GL} \ln^{-1/2} (w/ \lambda_\perp).
\end{equation}
In \Eqs{AL_IVC} and \eqref{AL_im}, $L$ is the film length, $I_c^{GL}$ is the GL
critical current formally calculated for the uniform current distribution and
$C$, $C'$ are fitting constants which cannot be determined within the framework
of the asymptotic approach.

\section{III. Results of numerical calculations}\vskip -3mm

In our calculations, we perform numerical solution of \Eq{eq1} with a certain
modification. As is obvious, the left-hand side of \Eq{eq1} is the approximate
form of \Eq{EqQ}, in which the vector potential $Q$ in the gradient term is
replaced by the current density $j$. This corresponds to the linear London
relation ${\bf j}\sim {\bf Q}$ between the current and the vector potential
which assumes independence of the order parameter of the vector potential. For
this reason, \Eq{eq1} is usually referred to as a generalized London equation
\cite{LO,AL}. This does not essentially affect the asymptotic results of
\cite{AL} because the gradient term is small at $w \gg \lambda_\perp$; however,
in our numerical calculations, we will use the full ``nonlinearized'' version
of \Eq{eq1} in a dimensionless form (see also \cite{BL}):
\begin{equation}
\label{eq3} \alpha \frac{dQ}{dx} + \frac{1}{4}V.p.\int_{-1}^1 \frac{i(x')dx'}
{x' - x} = - \frac{E'\,\textrm{sign}\,x}{i(x)},
\end{equation}
where the following definitions are introduced,
\begin{align}
\label{def1} &j(x) = \frac{3\sqrt {3}}{2} \frac{I_c^{GL}}{w} i(x), \quad i =
Q\left(1 - Q^2\right), \quad E = E'E_0,
\\
\label{def2} &E_0 = \frac{54\Phi_0}{\eta w^2 c^3}\bigl(I_c^{GL} \bigr)^2, \quad
\alpha = \frac{\pi \lambda_\bot}{2w}.
\end{align}

The distribution of the vector potential is obviously symmetric with respect to
the middle axis of the film, $Q(x) = Q(-x)$, which enables us to consider
\Eq{eq3} only in the region $x > 0$ and to reduce the integral in \Eq{eq3} to
the region of positive $x'$. After integration of the obtained equation from
the film edge to a given point $x$, we finally get
\begin{equation}
\label{eq4}
\alpha [ Q(x) - Q_e  ] = \frac{1}{4} \int_0^1 i(x')\ln\left| \frac{x^2 -
x'^2}{1 - x'^2} \right|dx' - E'\int_1^x \frac{dx'}{i(x')},
\end{equation}
where $Q_e \equiv Q(1)$ is the edge value of the vector potential. In these
notations, the net current $I$ is given by equation
\begin{equation}
\label{eq5} I = w\int_0^1 j(x)dx =  \frac{ 3\sqrt {3}}{2} I_c^{GL} \int_0^1
i(x)dx .
\end{equation}

\begin{figure}[b]
\centerline{\epsfxsize=8,7cm\epsffile{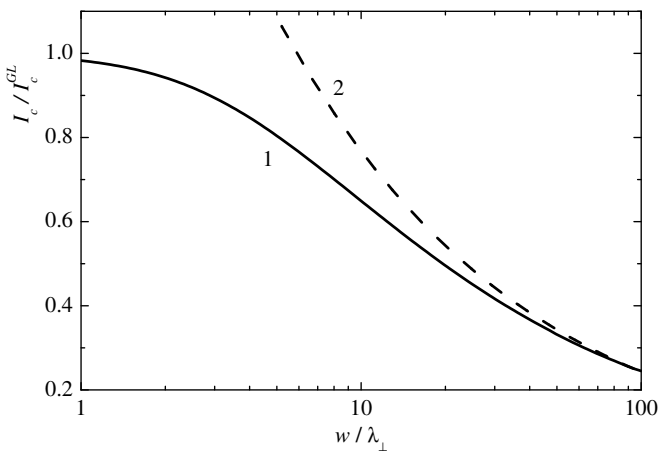}}\vskip -3mm
\caption{Result of numerical calculations of the critical current (curve 1) in
comparison with the asymptotical estimate \eqref{eq6} (curve 2).}
\label{fig1}\vskip -3mm
\end{figure}

At $I<I_c$, the quantity $Q_e$ increases with the current and has to be
determined self-consistently from \Eqs{eq4} and (\ref{eq5}) at zero electric
field; this procedure simultaneously gives the solution for the current
distribution across the film. As noted above, the resistive state of a wide
film at $I=I_c$ occurs when $Q_e$ approaches the critical value $Q_c = 1/
\sqrt{3}$. Such a relation is also obviously valid for narrow channels, that
makes it reasonable to extend it over the films of arbitrary width. In the
resistive vortex state, $I > I_c$, the quantity $Q_e$ holds its critical value
$Q_c$, and \Eqs{eq4} and (\ref{eq5}) determine the dependence $E'(I)$, i.e.,
the IVC, $V(I) = E'(I) E_0 L$.

A specific property of these equations is that their solutions, i.e., the
normalized current density distribution $i(x)$ and the electric field $E'$, are
universal functions of the parameters $w/\lambda_\bot$ and $I/I_c^{GL}$. This
implies that the normalized critical current $I_c/I_c^{GL}$ and the maximum
current of existence of the vortex state, $I_m/I_c^{GL}$, as well as the
normalized maximum electric field $E_m^{'}=E(I_m)/E_0$, are universal functions
of the parameter $w/\lambda_\bot$. Thus, the temperature dependencies of these
quantities, being expressed through the variable $w/\lambda_\bot(T)$, must
coincide for the films with different widths and thicknesses, which has been
demonstrated in experiments \cite{Dm1,Zol1}.

\subsection{A. Solution in subcritical regime $I \leqslant I_c$}\vskip -3mm

Solution of \Eqs{eq4} and (\ref{eq5}) can be found by an iteration method,
using $Q_e$ as the initial approximation for the function $Q(x)$. Although the
iteration parameter $\alpha^{-1} \sim w /\lambda_\bot$ is large for a wide
film, the convergence in this case can be nevertheless provided by introducing
certain weight factors for contributions of previous and current iterations.
The result of numerical calculation of the reduced critical current, shown in
Fig.~\ref{fig1}, describes transition from the uniformly distributed GL
depairing current $I_c^{GL} \propto \left(T_c - T \right)^{3/2}$ in a narrow
channel to the critical current $I_c^{AL} \sim T_c-T$ for a wide film
\eqref{eq6}. As seen from Fig.~\ref{fig1}, the asymptotic dependence
(\ref{eq6}) can be achieved with appropriate accuracy only at rather large
ratio $w/\lambda_\bot > 20-30$.

It should be noted that in some experiments \cite{Zol1,Dmitriev,Dm2}, the
behavior of $I_c(T)$ at the beginning of transition to the wide film regime was
found to be different from the smooth dependence following from the AL theory.
Namely, when the temperature decreases and the ratio $w/\lambda_\perp$ exceeds
$4\div5$, the critical current sharply falls to the value $I_c(T) \approx 0.8
I_c^{GL}(T)$ and holds this level until $w/\lambda_\perp \lesssim 10$. Within
this temperature interval, the film enters the vortex state at $I>I_c$,
although the temperature dependence of $I_c$ is similar to the case of a
vortex-free narrow channel. An analogous behavior of the critical current in
wide films has been registered in early experiments \cite{Meis1,Meis2}. To
explain such a specific dependence of $I_c(T)$, it was supposed in \cite{Zol1}
that the Pearl's vortices in moderately wide films may overcome the edge
barrier at the edge current density $\sim (1-T/T_c)^2$ much smaller than the GL
critical current density $\sim (1-T/T_c)^{3/2}$.

\begin{figure}[tb]
\centerline{\epsfxsize=8.5cm\epsffile{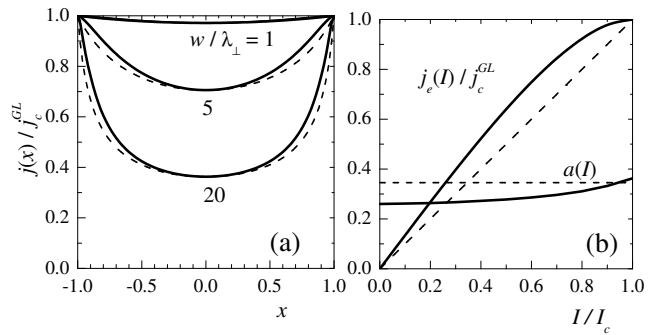}}\vskip -3mm
\caption{(a) - current distributions over the film width at the resistive
transition point, $I = I_c$, numerically calculated for different values
$w/\lambda_\bot = 1$, $5$, and $20$ (solid lines). Dashed lines show the
approximating dependence (\ref{eq7}) with the critical value of the coefficient
$a=a_c$ found from \Eq{eq8}; (b) - numerically calculated dependencies of the
edge current density $j_e$ and those of the current suppression coefficient in
the middle of the film, $a = j(0)/j_e$, on the transport current $I$ for the
wide film $w/\lambda_\bot = 20$ (solid lines). For comparison, the values
(\ref{eq9}) and (\ref{eq10}) found from the generalized London's equation
(\ref{eq1}) are shown by the dashed lines.}
\label{fig2}\vskip -3mm
\end{figure}

In Fig.~\ref{fig2}(a) we present the results of numerical calculations of the
current density distribution across the film at the resistive transition point
$I = I_c$ and different ratios $w/\lambda_\bot$. Similar results were obtained
in \cite{Deuk} by using London's equation for the superconducting current,
i.e., neglecting effect of the current on the order parameter, which is
equivalent to usage of \Eq{eq1}. As shown in \cite{Deuk}, a comparison of the
calculated and experimentally measured distributions of the supercurrent
density may be used for determination of the penetration depth $\lambda_\perp$.
Interestingly, these distributions are well approximated at arbitrary currents
by the function
\begin{equation}
\label{eq7}
j_1 (x) = j_e \frac{a}{\sqrt {1 - \left(1 - a^2\right)x^2}}.
\end{equation}
Equation \eqref{eq7} represents a modification of the asymptotic function
$j(x)=j(0) (1 - x^2)^{-1/2}$ in \cite{LO,Ivan,AL} with a regularization
parameter $a = j_1(0)/j_e$ which provides finiteness of the approximated
current density \eqref{eq7} at the film edges. As follows from its definition,
this parameter characterizes suppression of the current in the middle of the
film due to the Meissner screening. Substituting \Eq{eq7} with $j_e=j_c^{GL}
\equiv I_c^{GL}/w$ into \Eq{eq5}, we obtain equation for its value $a_c = \cos
\phi$ at the critical current,
\begin{equation}
\label{eq8}
{I_c}/{I_c^{GL}} = {\phi}/{\tan\phi}.
\end{equation}
In the case of a wide film, $w \gg \lambda_\bot$, the coefficient $a_c$ is
small, $a_c \ll 1$, and it can be estimated by using the asymptotic value
(\ref{eq6}) of the critical current as
\begin{equation}
\label{eq9} a_c = 2.74\left( \lambda_\bot/\pi w\right)^{1/2}.
\end{equation}

Within a framework of the generalized London's equation \eqref{eq1}, the
coefficient $a$ is independent of the current and holds a constant value $a_c$,
because in this approximation, the current distribution is determined only by
geometric factors and holds its shape at arbitrary currents $I < I_c$. The edge
current density in this case varies linearly with the transport current, that
reproduces the result of \cite{LO},
\begin{equation}
\label{eq10}
j_e = (I/I_c) j_c^{GL}
\end{equation}
[see dashed lines in Fig.~\ref{fig2}(b)]. Numerical calculations by means of
\Eq{eq3} demonstrate rather weak dependence $a(I)$ and nonlinearity of $j_e(I)$
shown in Fig. \ref{fig2}(b) by solid lines. In particular, the coefficient $a$
increases with the current and approaches a maximum value $a_c$ at $I = I_c$.
Physically, this is due to suppression of the order parameter by the transport
current, that weakens the screening effect while the current increases.

\subsection{B. Solution in the vortex state, $I_c < I \leqslant I_m$}\vskip -3mm

\begin{figure}[b]
\centerline{\epsfxsize=8.5cm\epsffile{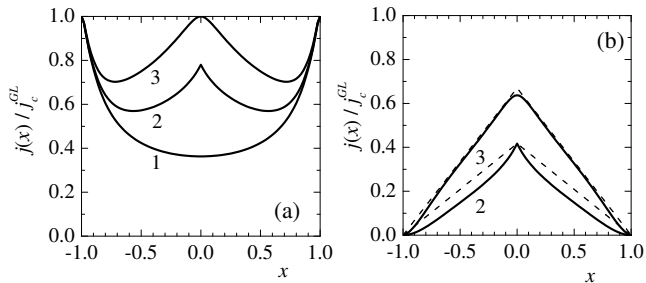}}\vskip -3mm
\caption{Solid lines - distributions of the net current density (a) and the
vortex contribution (b) in the resistive vortex state at $w/\lambda_\bot = 20$
numerically calculated at different values of the transport current: 1 - $I =
I_c$, 2 - $I = 0.5(I_c + I_m)$, 3 - $I = I_m$. The approximating distributions
(\ref{eq11}) are shown by dashed lines.}
\label{fig3}\vskip -3mm
\end{figure}
In the region of vortex resistivity, the distribution of the screening current
with sharp maxima of the height $j_c^{GL}$ at the film edges is superimposed by
the distribution associated with the vortex motion and having a peak at the
middle of the film, as shown in Fig.~\ref{fig3}. The logarithmic feature $\sim
{\ln^{1/2}(w/\lambda_\perp)}$ of this peak, predicted in \cite{AL}, is rather
weak and remains visible only for a certain intermediate current value; at $I
\to I_m$, this feature practically vanishes. In the experiment \cite{Siv}, an
inhomogeneous current distribution with three peaks in the vortex state of wide
films has been visualized by using the laser scanning microscope.

For moderately wide films, in which the above-mentioned logarithmic factor is
of the order of unity, the vortex contribution shown in Fig.~\ref{fig3}(b) can
be approximated by a piecewise-linear function
\begin{equation}
\label{eq11}
j_2 (x) = j_c^{GL} b( 1 - |x|)
\end{equation}
depicted in Fig.~\ref{fig3}(b) by dashed lines. As follows from \Eq{eq11}, the
parameter $b=j_2(0)/j_c^{GL}$ represents the relative (in units of $j_c^{GL}$)
current density created by vortices in the middle of the film. Within such an
approximation, this parameter, similar to the edge current density (\ref{eq10})
at $I < I_c$, linearly depends on the transport current,
\begin{equation}
\label{eq12}
b(I) = 2(I - I_c)/I_c^{GL}.
\end{equation}
\begin{figure}[tb]
\centerline{\epsfxsize=8.7cm\epsffile{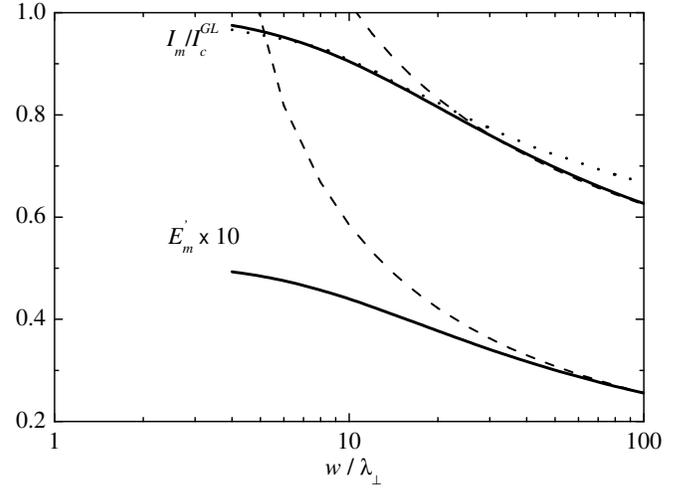}}\vskip -3mm
\caption{Dependencies of the maximum current of existence of the vortex state
$I_m$ and the normalized maximum electric field $E_m^{'}$ on the parameter
$w/\lambda_\bot$ (solid lines). Dashed lines show their asymptotic behavior
(\ref{eq13}) in the AL theory with the fitting constants $C_1 = 1.2$, $C_2 =
0.4$, $C_3 = 0.062$. Dotted line depicts the approximating dependence
\eqref{eq14} of $I_m$, in which the result of numerical calculation of $I_c$
(see Fig.~\ref{fig1}) and the formula \eqref{eq9} for the parameter $a_c$ were
used.}
\label{fig4}\vskip -3mm
\end{figure}

According to \cite{AL}, the vortex state becomes unstable when the height of
the central peak of the current distribution approaches the GL depairing
current density. Using this condition and solving \Eqs{eq3} and (\ref{eq4}) at
the critical edge value of the vector potential, $Q_e = Q_c$, we determine the
maximum current of existence of the vortex state $I_m$ and the normalized
maximum electric field $E_m^{'}=E'(I_m)$. The results of numerical calculation,
together with the asymptotical results of the AL theory, are presented in
Fig.~\ref{fig4}, in which the difference between the definitions of the
quantity $E_0$ introduced in our paper and in \cite{AL}, following from the
difference between \Eqs{eq1} and (\ref{eq3}), has been taken into account.

At large enough values of $w/\lambda_\bot \gtrsim 20-30$, the asymptotic
dependencies \cite{AL} shown in Fig.~\ref{fig4} by dashed lines,
\begin{equation}
\label{eq13}
I_m/I_c^{GL} = C_1 \ln^{ - 1/2}\left( C_2 w/\lambda_\bot\right), \quad E_m^{'}
= C_3 \left( I_m/I_c^{GL}\right)^2,
\end{equation}
can be fitted to the numerical results by an appropriate choice of the fitting
constants of the AL theory (shown in the caption of Fig.~\ref{fig4}) which
cannot be evaluated within the framework of the asymptotical analysis. Note
that in order to obtain a satisfactory agreement, one has to introduce an
additional constant $C_2$ into the argument of the logarithm, since the
formulas (\ref{eq13}) were derived in \cite{AL} within a logarithmic accuracy.
At smaller $w/\lambda_\perp \lesssim 20$, the asymptotic results \eqref{eq13}
considerably overestimate the values of $I_m$ and $E'_m$.

Another useful expression for $I_m$ suitable for a rather wide range of film
widths can be obtained from the approximating current distributions (\ref{eq7})
and (\ref{eq11}). At the stability threshold of the vortex state, where $j(0) =
j_1(0)+j_2(0)= j_c^{GL}$, the relation $b = 1 - a_c$ is fulfilled which leads
to equation
\begin{equation}
\label{eq14}
I_m = I_c + 0.5I_c^{GL} (1 - a_c).
\end{equation}
As seen from Fig.~\ref{fig4}, this approximation (dotted line) rather well
reproduces the result of numerical calculations of $I_m$, up to the point of
nucleation of the vortex resistivity at $w/\lambda_\perp \approx 4$. Of course,
for extremely wide films, in which $I_m \gg I_c$ and the logarithmic peak of
the current is well pronounced, the AL asymptotic expression \eqref{eq13} for
$I_m$ with numerically calculated fitting constants is more preferable.

\begin{figure}[tb]
\centerline{\epsfxsize=7.5cm\epsffile{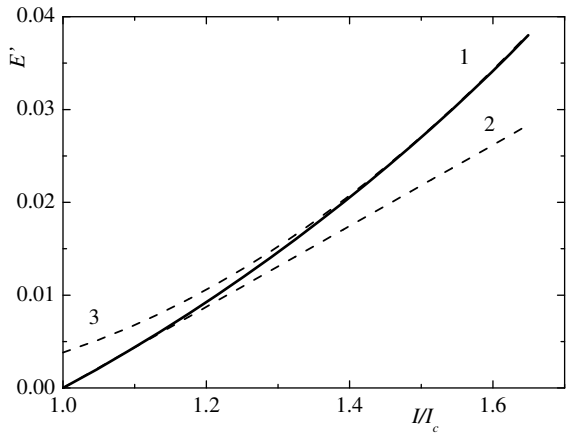}}\vskip -3mm
\caption{Numerically calculated IVC of the superconducting film in the vortex
state at $w/\lambda_\bot = 20$ (solid line 1). Dashed straight line 2 is the
linear AL asymptotics for $I - I_c \ll I_c$ (\ref{eq15}), and the parabola 3 is
the shifted AL asymptotics for $I \gg I_c $ (\ref{eq16}). }
\label{fig5}\vskip -3mm
\end{figure}

In Fig.~\ref{fig5}, the calculated normalized IVC per unit length of a wide
film ($w/\lambda_\bot = 20$) is shown by the curve 1 in the region of existence
of the stable vortex state $I_c < I < I_m \approx 1.7I_c$. Its initial part
coincides with the linear AL asymptotics (line 2) at $I - I_c \ll I_c$, which
is described by the formula
\begin{equation}
\label{eq15}
E'(I) = \frac{4}{27}\Bigl( \frac{I_c}{I_c^{GL}}\Bigr)^2\Bigl( \frac{I}{I_c} - 1
\Bigr) \approx 0.873\frac{\lambda_\bot}{w} \Bigl(\frac{I}{I_c} - 1 \Bigr),
\end{equation}
obtained by using approximate \Eq{eq6} for $I_c/I_c^{GL}$. At $I
> 1.4I_c $, the calculated characteristic is well described by the modified AL
asymptotics for $I \gg I_c$:
\begin{equation}
\label{eq16}
E'(I) = C_1 \frac{4}{27} \Bigl( \frac{I_c}{I_c^{GL}} \Bigr)^2 \Bigl(
\frac{I}{I_c} - C_2 \Bigr)^2 \approx 0.873 C_1 \frac{\lambda_\bot}{w}\Bigl(
\frac{I}{I_c} - C_2\Bigr)^2
\end{equation}
with the fitting constants $C_{1} = 0.97$ and $C_{2} = 0.7$. Introduction of an
additional constant $C_2$, which shifts the original AL parabola, enables us to
extrapolate the result obtained in \cite{AL} for the case of large
supercriticality, $I \gg I_c$, to the region of currents comparable with $I_c$.
Such a modification of the AL asymptotic formulas has been successfully used
for fitting of the parabolic part of the IVC in \cite{Zol2}. In experiments
with relatively narrow films (in which the vortex state nevertheless exists),
only a linear part of the IVC is observed, because the region of the vortex
resistivity is rather narrow in this case.

In conclusion to this Section, we note that a numerical simulation of the
vortex motion in an infinitely long and thick superconducting slab \cite{Vod}
gives similar results for the current distribution and the IVCs, although these
results cannot be directly applied to the thin film because of essential
difference between strongly localized Abrikosov vortices in a bulk slab and
Pearl vortices in a thin film which interact mostly via the fields in the
surrounding space \cite{Pearl,Kogan}.

\section{IV. Summary}\vskip -3mm
%%%%%%%%%%%%%%%%%%%%%%%%%%%%%

We studied the distributions of the transport current in thin superconducting
films in zero external magnetic field within a wide range of the film widths
$w$ and temperatures, using numerical solutions of the integro-differential
equations for the gauge-invariant vector potential . These distributions can be
effectively approximated by rather simple analytical formulas, the parameters
of which have a clear physical sense and can be relatively easily calculated.

We calculated universal dependencies of the critical current $I_c$ and the
maximum current of existence of the vortex state $I_m$ (normalized on the
Ginzburg-Landau critical current in a uniform current state), as well as the
dependencies of the reduced maximum electric field in the vortex state on the
parameter $w/\lambda_\perp$. For wide enough films, $w/\lambda_\perp \gtrsim
20-30$, our numerical results coincide with the asymptotical dependencies found
in \cite{AL,BL}. We study numerically the current-voltage characteristic of a
wide film in the vortex state and propose a modification of the asymptotical
results of \cite{AL} which provides much better fitting with the experimental
data. The pinning of the vortices can be also taken into account \cite{AL} by a
certain modification of \Eq{v} and will be considered elsewhere.

In conclusion, we note that the validity of our results is confined by the
boundaries of applicability of the static Ginzburg-Landau equations to the
solution of the problem under consideration. Within this approach, the order
parameter relaxation time $\tau_\Delta$ is assumed to be much smaller than
other characteristic times of the system. In a general case, the finiteness of
$\tau_\Delta$ results in deformation of the vortex core and in occurrence of a
wake with the suppressed order parameter behind the moving vortex. As shown by
numerical simulation \cite{Vod}, this may anomalously enhance the vortex
velocity and lead to creation of fast moving chains of vortices treated in
\cite{Vod} as the nuclei of the phase-slip lines. The instability of the vortex
motion due to the nonequilibrium state of quasiparticles in the vortex core
\cite{LO1} is also neglected, although it can be taken into account
phenomenologically by introducing the dependence of the viscosity coefficient
$\eta$ in \eqref{v} on the vortex velocity. Finally, the AL model assumes
rather weak pinning of the vortices, the penetration of which into the film is
followed by their continuous viscous motion leading to the current dissipation.
The opposite case of strong pinning corresponds to the model of the critical
state with unmovable vortices, which results in quite different distributions
of the current and magnetic field (see, e.g., \cite{Crit}).

The author is grateful to I.V.~Zolochevskii for helpful discussions and
advices.


\begin{thebibliography}{99}
\bibitem{footnote} Actually, according to the definition of $Q$, the correct
formula for the dimensionless current density must have the opposite sign, $j =
-Q(1-Q^2)$. Thus, to deal with positive values of the transport current, one
should consider negative values of $Q$. To avoid this inconvenience, the
authors of \cite{LO,AL} define $j$ as presented in the text and use positive
$Q$ values, that does not affect the results. In our paper, we also follow this
convention.
\end{thebibliography}
\end{document}